\documentclass[aps,preprint,groupedaddress,superscriptaddress]{revtex4-1}

\usepackage{amsmath}
\usepackage{amssymb}
\usepackage{graphicx}
\usepackage{subfigure}
\usepackage{dcolumn}
\usepackage{array}
\usepackage{lineno}
\usepackage{bm}
\usepackage{color}
\usepackage[colorlinks,linkcolor=blue, urlcolor=blue, anchorcolor=blue, citecolor=blue]{hyperref}

\begin{document}

\title{Analysis of $\mathit{CP}$ violation in $D^0 \to K^+K^-\pi^0$}

\author{Hang Zhou
}
\email{hang\_zhou@outlook.com}
\affiliation{School of Nuclear Science and Technology, University of South China, Hengyang, Hunan 421001, China}
\author{Bo Zheng}
\email{Corresponding author, zhengbo\_usc@163.com}
\affiliation{School of Nuclear Science and Technology, University of South China, Hengyang, Hunan 421001, China}
\affiliation{Helmholtz Institute Mainz, Johann-Joachim-Becher-Weg 45, D-55099 Mainz, Germany}
\author{Zhen-Hua Zhang
}
\email{Corresponding author, zhangzh@usc.edu.cn}

\affiliation{School of Nuclear Science and Technology, University of South China, Hengyang, Hunan 421001, China}

\date{\today}
\begin{abstract}

We study the $\mathit{CP}$ violation  induced by the interference between two intermediate resonances $K^*(892)^+$ and $K^*(892)^-$ in the phase space of singly-Cabibbo-suppressed decay $D^0 \to K^+K^-\pi^0$. We adopt the factorization-assisted topological approach in dealing with the decay amplitudes of $D^0 \to K^\pm K^*(892)^\mp$. The $\mathit{CP}$ asymmetries of two-body decays are predicted to be very tiny, which are $(-1.27 \pm 0.25) \times 10^{-5}$ and $(3.86 \pm 0.26) \times 10^{-5}$ respectively for $D^0 \to  K^+ K^*(892)^-$ and $D^0 \to  K^- K^*(892)^+$.
While the differential $\mathit{CP}$ asymmetry of $D^0 \to K^+K^-\pi^0$ is enhanced because of the interference between the two intermediate resonances, which can reach as large as $3\times10^{-4}$. For some NPs which have considerable impacts on the chromomagnetic dipole operator $O_{8g}$, the global $\mathit{CP}$ asymmetries of $D^0 \to K^+ K^*(892)^- $ and $D^0 \to K^- K^*(892)^+ $ can be then increased to $(0.56\pm0.08)\times10^{-3}$ and $(-0.50\pm0.04)\times10^{-3}$, respectively. The regional $\mathit{CP}$ asymmetry in the overlapped region of the phase space can be as large as $( 1.3\pm 0.3)\times10^{-3}$.

\end{abstract}


\maketitle

\section{introduction\label{sec:intro}}

Charge-Parity ($\mathit{CP}$) violation, which was first discovered in $K$ meson system in 1964~\cite{Christenson:1964fg},  is one of the most important phenomenon in particle physics.
In the Standard Model (SM), $\mathit{CP}$ violation originates from the weak phase in the Cabibbo-Kobayashi-Maskawa (CKM) matrix~\cite{Kobayashi:1973fv,Cabibbo:1963yz}, and the unitary phases which usually arise from strong interactions. 
One reason for the smallness of $\mathit{CP}$ violation is that the unitary phase is usually small. 
Nevertheless, $\mathit{CP}$ violation can be enhanced in three-body decays of heavy hadrons, when the corresponding decay amplitudes are dominated by overlapped intermediate resonances in certain regions of phase space.
Owing to the overlapping, a regional $\mathit{CP}$ asymmetry can be generated by a relative strong phase between amplitudes corresponding to different resonances.
This relative strong phase has non-perturbative origin. 
As a result, the regional $\mathit{CP}$ asymmetry can be larger than the global one.
In fact, such kind of enhanced $\mathit{CP}$ violation has been observed in several three-body decay channels of  $B$ meson~\cite{Aaij:2013sfa,Aaij:2013bla,Aaij:2014iva,Amato:2016xjv},
which was followed by a number of theoretical works~\cite{Zhang:2013oqa,Bediaga:2013ela,Cheng:2013dua,Zhang:2013iga,Bhattacharya:2013boa,Xu:2013dta,Wang:2014ira,Zhang:2015wba,Wang:2015ula,Nogueira:2015tsa,Dedonder:2010fg,ElBennich:2006yi}.

The study of $\mathit{CP}$ violation in singly-Cabibbo-suppressed (SCS) $D$ meson decays provides an ideal test of the SM and exploration of New Physics (NP)~\cite{Bigi:1986dp,Blaylock:1995ay,Bergmann:2000id,Nierste:2017cua}.
In the SM,  $\mathit{CP}$ violation is predicted to be very small in charm system. Experimental researches have shown that there is no significant $\mathit{CP}$ violation so far in charmed hadron decays~\cite{Bonvicini:2000qm,Link:2001zj,Aaltonen:2011se,Cenci:2012ru,Lees:2012jv,Staric:2015sta,Aaij:2017eux,Aaij:2016roz,Aaij:2016nki,Bhardwaj:2017hsw}. 
$\mathit{CP}$ asymmetry in SCS $D$ meson decay can be as small as 
\begin{equation}
A_\mathit{CP}\sim\frac{|V_{cb}^*V_{ub}|}{|V_{cs}^*V_{us}|}\frac{\alpha_s}{\pi}\sim10^{-4},
\end{equation}
or even less, due to the suppression of the penguin diagrams by the CKM matrix as well as the smallness of Wilson coefficients in penguin amplitudes.
The SCS decays are sensitive to new contributions to the $\Delta C=1$ QCD penguin and chromomagnetic dipole operators, while such contributions can affect neither the Cabibbo-favored (CF) $(c \to s \bar d u)$ nor the doubly-Cabibbo-suppressed (DCS) $(c \to d \bar s u)$ decays~\cite{Grossman:2006jg}.
Besides, the decays of charmed mesons offer a unique opportunity to probe $\mathit{CP}$ violation in the up-type quark sector.

Several factorization approaches have been wildly used in non-leptonic $B$ decays. In the naive factorization approach ~\cite{Bjorken:1988kk,Dugan:1990de}, the hadronic matrix elements were expressed as a product of a heavy to light transition form factor and a decay constant. Based on Heavy Quark Effect Theory, it is shown in  the QCD  factorization approach that the corrections to the hadronic matrix elements can be expressed in terms of short-distance coefficients and meson light-cone distribution amplitudes~\cite{Beneke:1999br,Beneke:2003zv}. Alternative factorization approach based on QCD factorization is often applied in study of quasi two-body hadronic $B$ decays~\cite{Boito:2017jav,Furman:2005xp,ElBennich:2006yi}, where they introduced unitary meson-meson form factors, from the perspective of unitarity, for the final state interactions.  
Other QCD-inspired approaches, such as the perturbative approach (pQCD)~\cite{Keum:2000wi} and the soft-collinear effective theory (SCET)~\cite{Bauer:2004tj}, are also wildly used in $B$ meson decays. 

However, for $D$ meson decays, such QCD-inspired factorization approaches may not reliable since the charm quark mass, which is just above 1 GeV, is not heavy enough for the heavy quark expansion~\cite{Loiseau:2016mdm,Cheng:2012wr}.
For this reason, several model-independent approaches for the charm meson decay amplitudes have been proposed, such as the flavor topological diagram approach based on the flavor $SU(3)$ symmetry~\cite{Chau:1982da,Cheng:2012wr,Bhattacharya:2012ah,Cheng:2016ejf}, and the factorization-assisted topological-amplitude (FAT) approach with the inclusion of flavor $SU(3)$ breaking effect~\cite{Li:2012cfa,Li:2013xsa}.
One motivation of these aforementioned approaches is to identify as complete as possible the dominant sources of non-perturbative dynamics in the hadronic matrix elements.

In this paper, we study the $\mathit{CP}$ violation of SCS $D$ meson decay $D^{0} \to K^{+}K^{-}\pi^{0}$ in the FAT approach. Our attention will be mainly focused on the region of the phase space where two intermediate resonances, $K^{*}(892)^{+}$ and $K^{*}(892)^{-}$, are overlapped. Before proceeding, it will be helpful to point out that direct $ \mathit{CP} $ asymmetry is hard to be isolated for decay process with $ \mathit{CP} $-eigen-final-state.
When the final state of the decay process is $\mathit{CP}$ eigenstate, the time integrated $\mathit{CP}$ violation for $ D^0\to f $, which is defined as
\begin{equation}
a_f\equiv\frac{\int^\infty_0\Gamma(D^0\to f)dt-\int^\infty_0\Gamma(\bar{D}^0\to f)dt}{\int^\infty_0\Gamma(D^0\to f)dt+\int^\infty_0\Gamma(\bar{D}^0\to f)dt},
\end{equation}
can be expressed as~\cite{Grossman:2006jg},
\begin{equation}
a_f=a_f^d+a_f^m+a_f^i,
\end{equation}
where $a_f^d$, $a_f^m$, and $a_f^i$, are the $\mathit{CP}$ asymmetries in decay, in mixing, and in the interference of decay and mixing, respectively.
As is shown in Ref.~\cite{Grossman:2006jg,Aubert:2003pz,Abe:2003ys}, the indirect $\mathit{CP}$ violation $a^\mathrm{ind}\equiv a^m+a^i$ is universal and  channel-independent for two-body $\mathit{CP}$-eigenstate.
This conclusion is easy to be generalized to decay processes with three-body $\mathit{CP}$-eigenstate in the final state, such as $D^0\to K^+K^-\pi^0$.
In view of the universality of the indirect $\mathit{CP}$ asymmetry, we will only consider the direct $\mathit{CP}$ violations of the decay $D^0\to K^+K^-\pi^0$ throughout this paper.

The remainder of this paper is organized as follows. 
In Section~\ref{sec:3b}, we present the decay amplitudes for various decay channels, where, the decay amplitudes of $D^0 \to K^{\pm}K^*(892)^{\mp}$ are formulated via the FAT approaches. 
In Section~\ref{sec:CP}, we study the $\mathit{CP}$ asymmetries of $D^0 \to K^{\pm}K^*(892)^{\mp}$ and the $\mathit{CP}$ asymmetry of $D^0\to K^+K^-\pi^0$ induced by the interference between different resonances in the phase space. 
Discussions and conclusions are given in Section~\ref{sec:dis-con}. 
We list some useful formulas and input parameters in the Appendix~\ref{app:input}.

%%%%%%%%%%
%%%%%%%%%%
\section{Decay amplitude for {\boldmath$D^0 \to K^+K^-\pi^0$}\label{sec:3b}}
%%%%%%%%%%
%%%%%%%%%%
In the overlapped region of the intermediate resonances $K^*(892)^+$ and $K^*(892)^-$ in the phase space, the decay process $D^0  \to K^+K^-\pi^0$ is dominated by two cascade decays, $D^0  \to K^+K^*(892)^- \to K^+K^-\pi^0$ and $D^0  \to K^-K^*(892)^+ \to K^-K^+\pi^0$, respectively.
Consequently, the decay amplitude of $D^0  \to K^+K^-\pi^0$ can be express as
\begin{equation}
\mathcal{M}_{D^0 \to K^+K^-\pi^0}=\mathcal{M}_{K^{*+}}+e^{i\delta}\mathcal{M}_{K^{*-}} \label{eq:totampl}
\end{equation}
in the overlapped region, where $\mathcal{M}_{K^{*+}}$ and $\mathcal{M}_{K^{*-}}$ are the amplitudes for the two cascade decays, and $\delta$ is the relative strong phase.
Note that non-resonance contributions have been neglect in Eq.~(\ref{eq:totampl}).

The decay amplitude for the cascade decay $D^0  \to K^+K^*(892)^- \to K^+K^-\pi^0$ can be expressed as
\begin{equation}
\mathcal{M}_{K^{*-}}=\frac{\sum_\lambda\mathcal{M}^\lambda_{K^{*-}\to K^-\pi^0}\cdot\mathcal{M}^\lambda_{D^0\to K^{*-}K^+}}{s_{\pi^0K^-}-m_{K^{*-}}^2+im_{K^{*-}}\Gamma_{K^{*-}}},\label{eq:oneampl}
\end{equation}
where $\mathcal{M}^\lambda_{K^{*-}\to K^-\pi^0}$ and $\mathcal{M}^\lambda_{D^0\to K^+K^{*-}}$ represent the amplitudes corresponding to the strong decay $K^{*-}\to K^-\pi^0$ and weak decay $D^0\to K^+K^{*-}$, respectively, $\lambda$ is the helicity index of $ K^{*-} $, $s_{\pi^0K^-}$ is the invariant mass square of $\pi^0K^-$ system, $m_{K^{*-}}$ and $\Gamma_{K^{*-}}$ are the mass and width of $K^*(892)^-$, respectively.
The decay amplitude for the cascade decay, $D^0  \to K^- K^*(892)^+ \to K^-K^+\pi^0$, is the same as Eq.~(\ref{eq:oneampl}) except replacing the subscript $K^{*-}$ and $K^\pm$ with $K^{*+}$ and $K^\mp$, respectively.

For the strong decays $K^*(892)^\pm \to \pi^0K^\pm$, one can express the decay amplitudes as
\begin{equation}
\mathcal{M}_{K^{*\pm}\to \pi^0 K^\pm}=g_{K^{*\pm}K^\pm \pi^0}(p_{\pi^0}-p_{K^\pm})\cdot\varepsilon_{K^{*\pm}}(p,\lambda),
\end{equation}
where $p_{\pi^0}$ and $p_{K^\pm}$ represent the momentum for $\pi^0$, $K^\pm$ mesons, respectively, $g_{K^{*\pm}K^\pm \pi^0}$ is the effective coupling constant for the strong interaction, which can be extracted from the experimental data via
\begin{equation}
g_{K^{*\pm}K^\pm \pi^0}^2=\frac{6\pi m_{K^{*\pm}}^2\Gamma_{K^{*\pm} \to K^\pm \pi^0}}{\lambda_{K^{*\pm}}^3},
\end{equation}
with
\begin{equation}
\lambda_{K^{*\pm}}=\frac{1}{2m_{K^{*\pm}}}\sqrt{\left[m_{K^{*\pm}}^2-(m_{\pi^0}+m_{K^\pm})^2\right]\cdot\left[m_{K^{*\pm}}^2-(m_{\pi^0}-m_{K^\pm})^2\right]},
\end{equation}
and $\Gamma_{K^{*\pm}\to K^\pm \pi^0}=\mathrm{Br}(K^{*\pm}\to K^\pm \pi^0)\cdot\Gamma_{K^{*\pm}}$. 
The isospin symmetry of the strong interaction implies that $\Gamma_{K^{*\pm} \to K^\pm \pi^0}\simeq\frac{1}{3}\Gamma_{K^{*\pm}}$.

\begin{figure}[h]
	\centering
	\subfigure[\ $T_{P(V)}$]{\label{fig:T}
		\includegraphics[width=5cm]{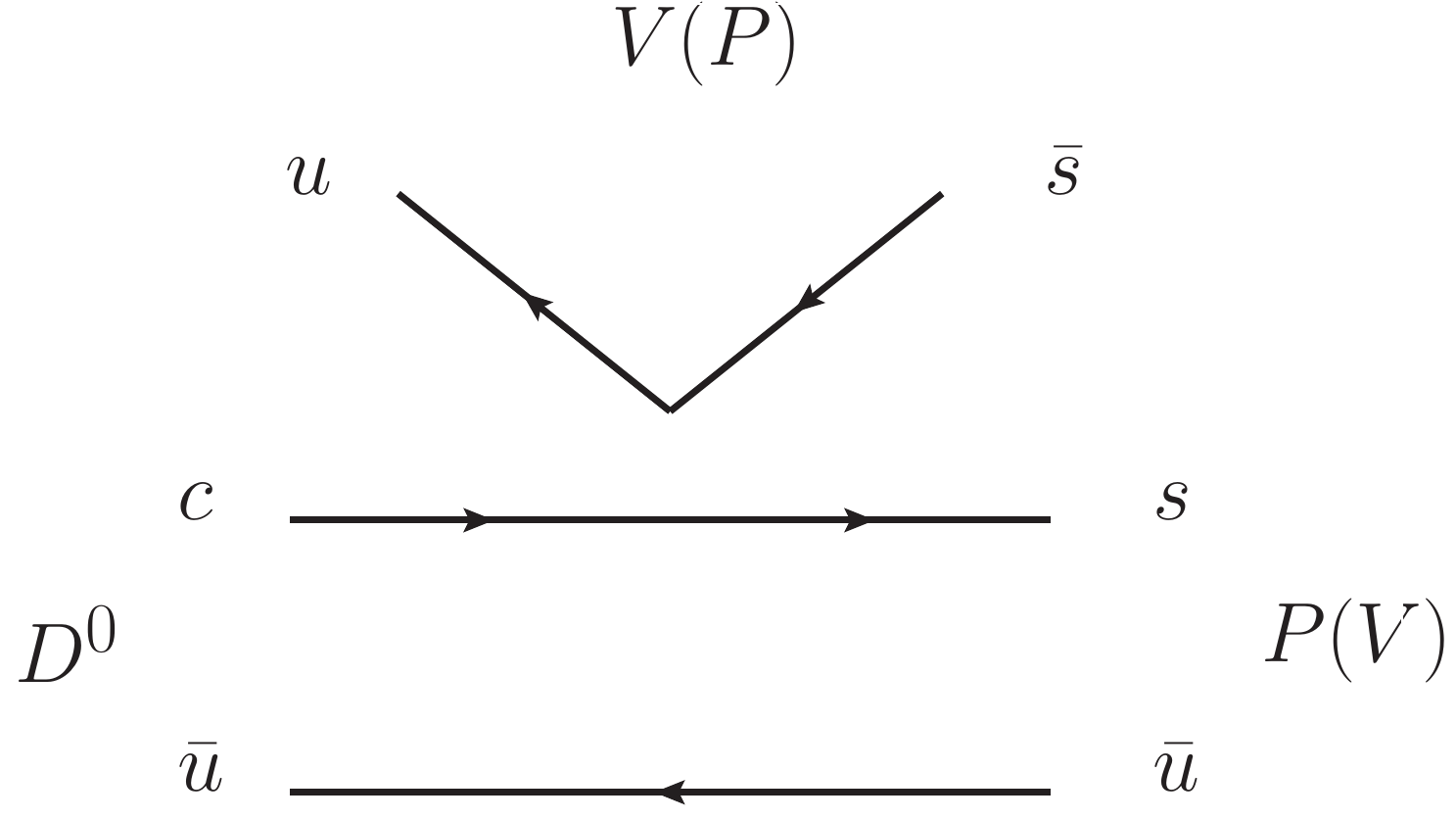}}
	\hspace{0.5in}
	\subfigure[\ $E_{P(V)}$]{\label{fig:E}
		\includegraphics[width=5cm]{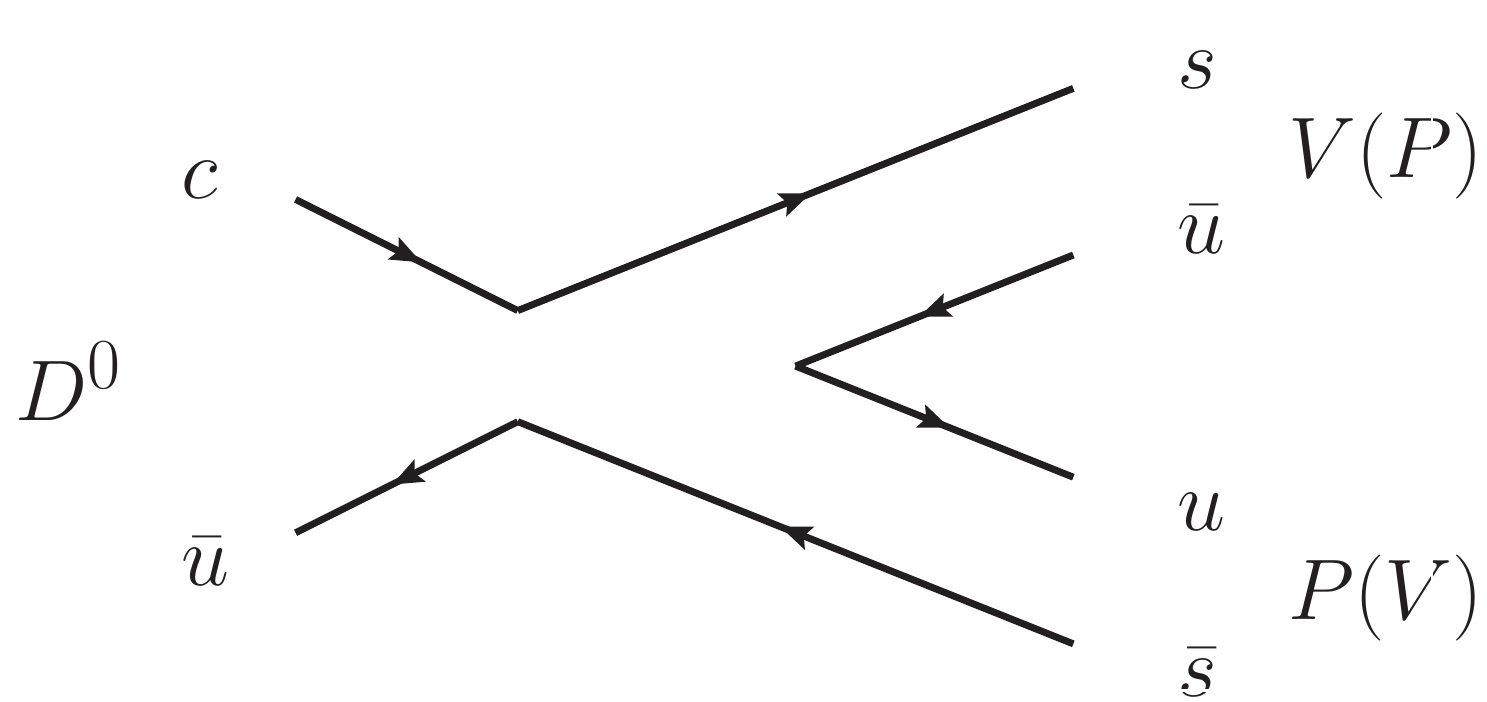}}
	\hspace{5cm}
	\subfigure[\ $PT_{P(V)}$]{
		\label{fig:PT}
		\includegraphics[width=5cm]{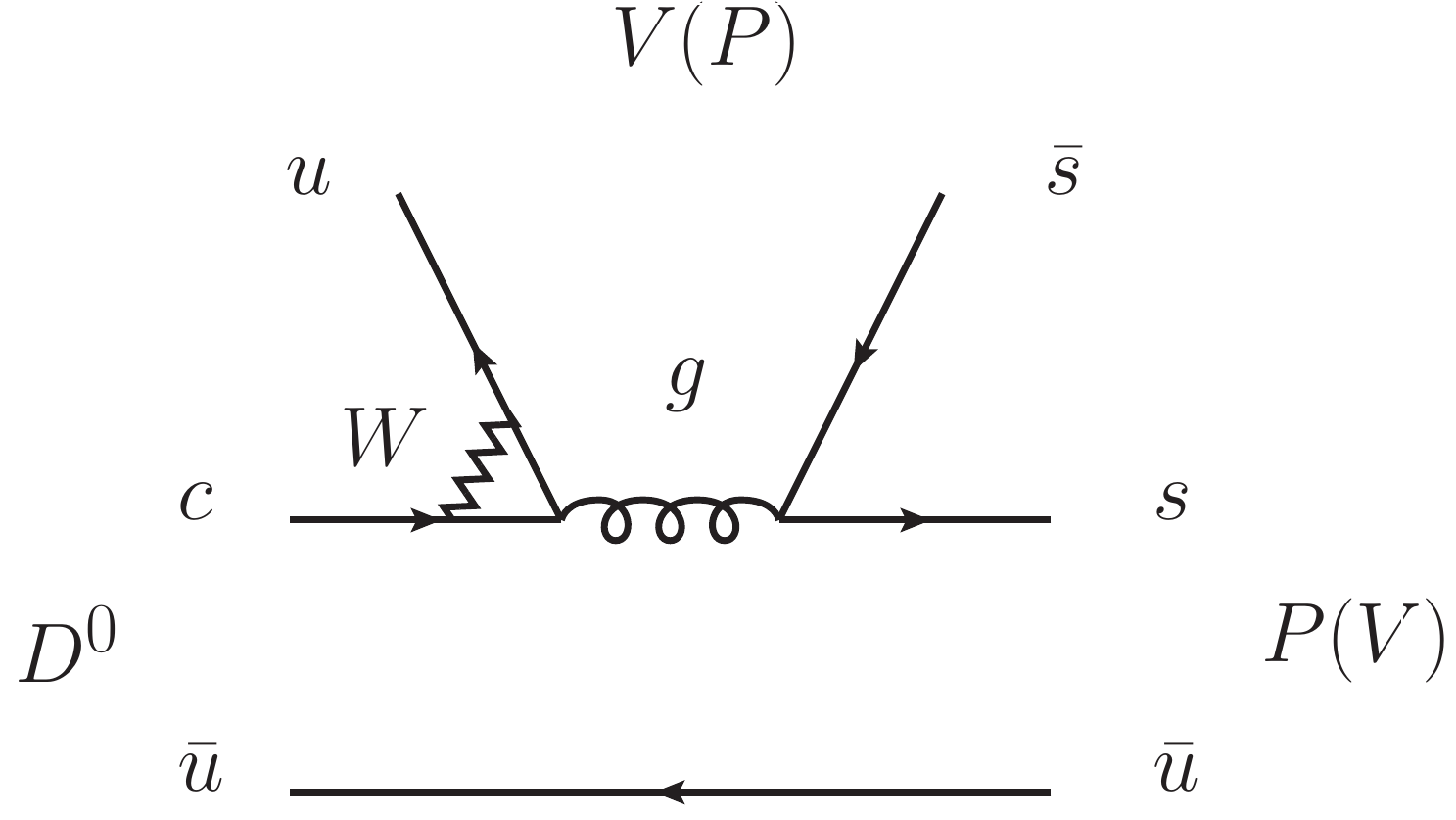}}
	\hspace{0.1in}
	\subfigure[\ $PE_{P(V)}$]{
		\label{fig:PE}
		\includegraphics[width=5cm]{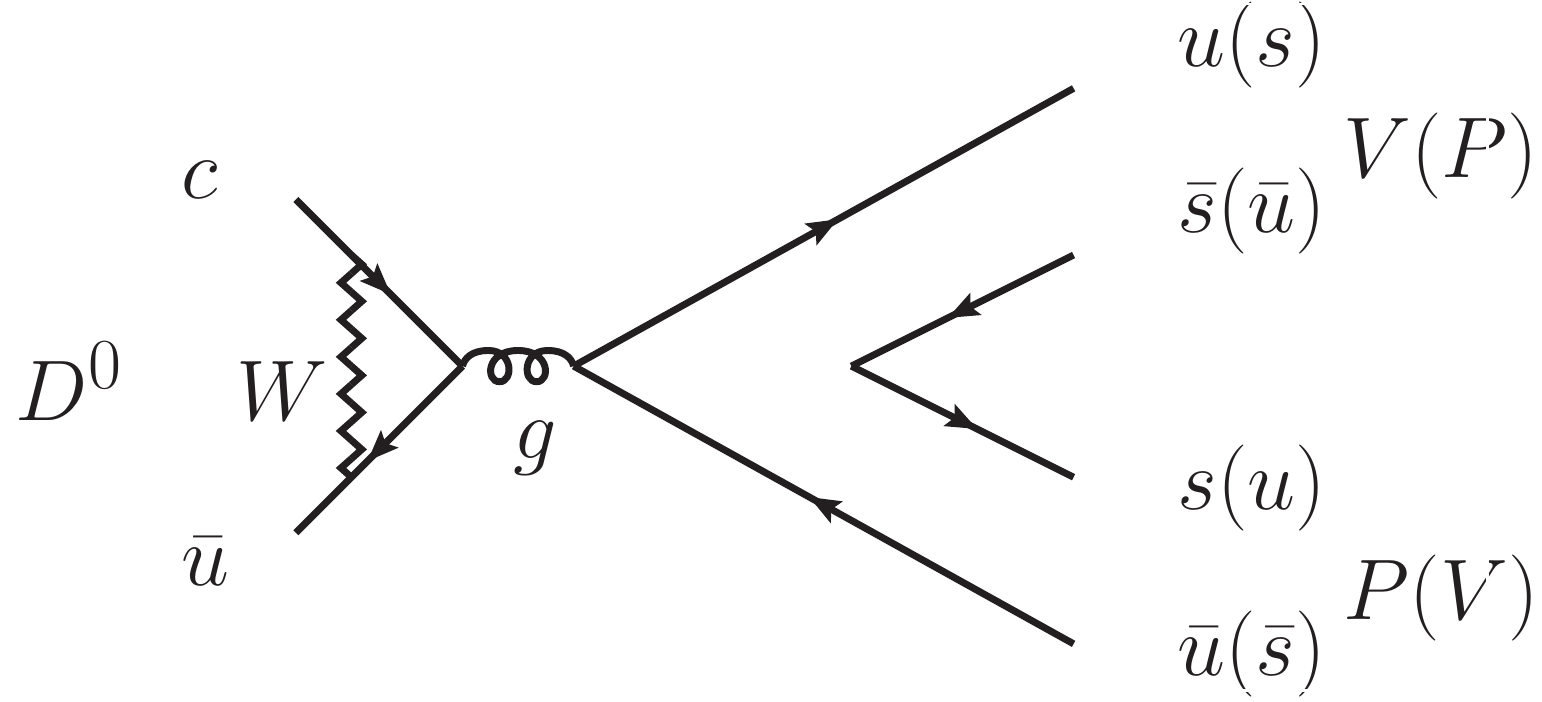}}
	\hspace{0.1in}
	\subfigure[\ $PA_{P(V)}$]{
		\label{fig:PA}
		\includegraphics[width=5cm]{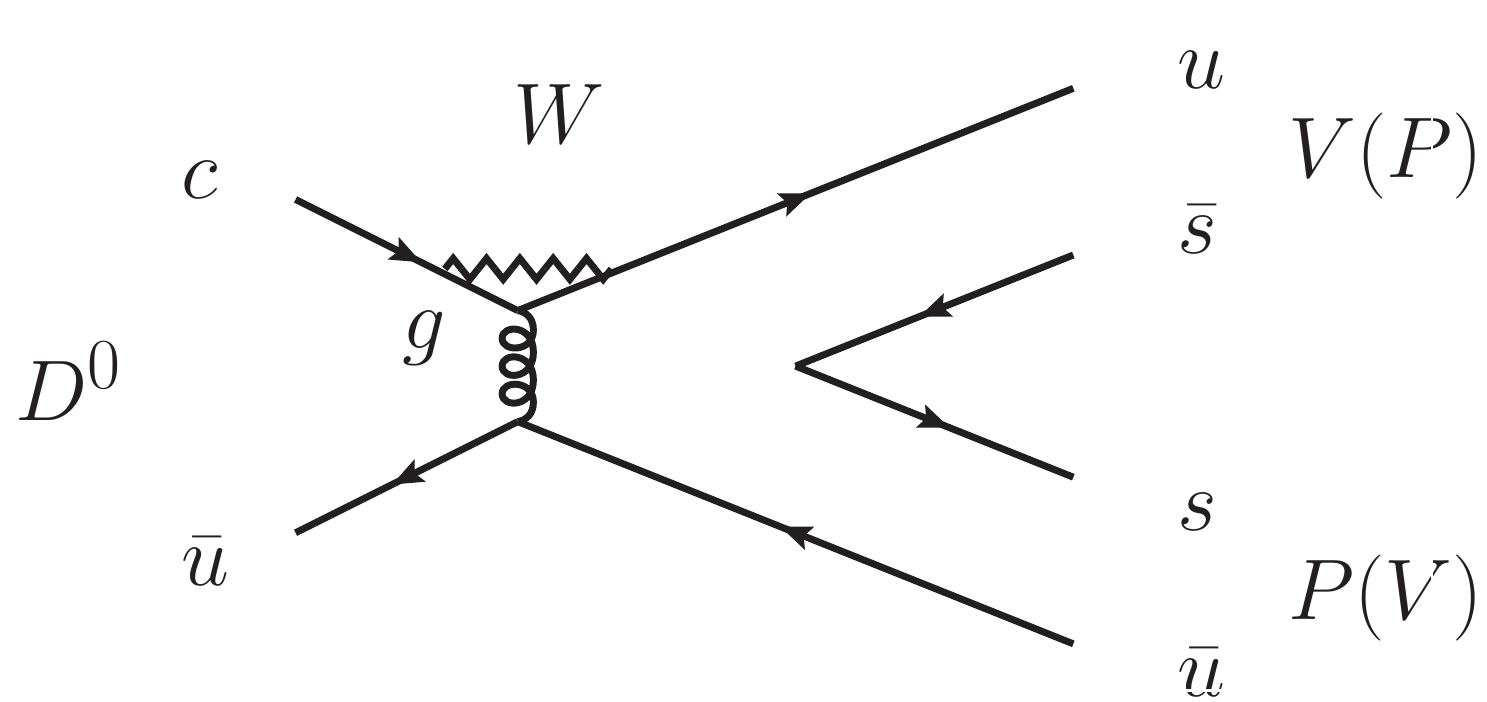}}
	\caption{The relevant topological diagrams for $D \to PV$ with  (a) the color-favored tree amplitude $T_{P(V)}$, (b) the $W$-exchange amplitude $E_{P(V)}$, (c) the color-favored penguin amplitude $PT_{P(V)}$, (d) the gluon-annihilation penguin amplitude $PE_{P(V)}$, and (e) the gluon-exchange penguin amplitude $PA_{P(V)}$.}
	\label{fig:topod}
\end{figure}

The decay amplitudes for the weak decays, $D^0\to K^+K^{\ast}(892)^{-}$ and $D^0\to K^-K^{\ast}(892)^{+}$, will be handled with the aforementioned FAT approach ~\cite{Li:2012cfa,Li:2013xsa}. The relevant topological tree and penguin diagrams for $D \to PV$ are displayed in Fig.~\ref{fig:topod}, where $P$ and $V$ denotes a light pseudoscalar and vector meson (representing $K^\pm$ and $K^{*\pm}$ in this paper), respectively.

The two tree diagrams in first line of Fig.~\ref{fig:topod} represent the color-favored tree diagram for $D \to P(V)$ transition, and the $W$-exchange diagram with the pseudoscalar (vector) meson containing the anti-quark from the weak vertex, respectively.
The amplitudes of these two diagrams will be respectively denoted as $T_{P(V)}$ and $E_{P(V)}$.

According to these topological structure, the amplitudes of the color-favored tree diagrams $T_{P(V)}$, which is dominated by the factorizable contributions, can be parameterized as
\begin{equation}
T_P=\frac{G_F}{\sqrt{2}}\lambda_sa_2(\mu)f_Vm_VF_1^{D\to P}(m_V^2)2(\varepsilon^*\cdot p_D),
\end{equation}
and
\begin{equation}
T_V=\frac{G_F}{\sqrt{2}}\lambda_sa_2(\mu)f_Pm_VA_0^{D\to V}(m_P^2)2(\varepsilon^*\cdot p_D),
\end{equation}
respectively, where $G_F$ is the Fermi constant, $\lambda_s=V_{us}V_{cs}^*$, with $V_{us}$ and $V_{cs}$ being the CKM matrix elements, $a_2(\mu)=c_2(\mu)+c_1(\mu)/N_c$, with $c_1(\mu)$ and $c_2(\mu)$ being the scale-dependent Wilson coefficients, and the number of color $N_c=3$,
$f_{V(P)}$ and $m_{V(P)}$ are the decay constant and mass of the vector (pseudoscalar) meson, respectively, $F_1^{D\to P}$ and $A_0^{D\to V}$ are the form factors for the transitions $D\to P $ and $D\to V$, respectively, $\varepsilon$ is the polarization vector of the vector meson, and $p_D$ is the momentum of $D$ meson.  
The scale $\mu$ of Wilson coefficients is set to energy release in individual decay channels~\cite{Keum:2000ph,Lu:2000em}, which depends on masses of initial and final states, 
and is defined as~\cite{Li:2012cfa,Li:2013xsa} 
\begin{equation}
\mu=\sqrt{\Lambda m_D (1-r_P^2)(1-r_V^2)},
\end{equation} 
with the mass ratios $r_{V(P)}=m_{V(P)}/m_D$, where $\Lambda$ represents the soft degrees of freedom in the $D$ meson, which is a free parameter.

For the $W$-exchange amplitudes, since the factorizable contributions to these amplitudes are helicity-suppressed, only the non-factorizable contributions need to be considered. Therefore, the $W$-exchange amplitudes are parameterized as
\begin{equation}
E^q_{P,V}=\frac{G_F}{\sqrt2}\lambda_sc_2(\mu)\chi_{q}^Ee^{i\phi_{q}^E}f_Dm_D\frac{f_Pf_V}{f_\pi f_\rho}(\varepsilon^*\cdot p_D),
\end{equation}
where $m_D$ is the mass of $D$ meson, $f_D$, $f_\pi$ and $f_\rho$ are the decay constants of the $D$, $\pi$, and $\rho$ mesons, respectively, $\chi_{q}^E$ and $\phi_{q}^E$ characterize the strengths and the strong phases of the corresponding amplitudes, with $q=u, d, s$ representing the strongly produced $q$ quark pair. The ratio of $f_Pf_V$ over $f_\pi f_\rho$ indicates that the flavor $SU(3)$ breaking effects have been taken into account from the decay constants. 

%%%%%%%%
%%%%%%%%
%%%%%%%%
%%%%%%%%

The penguin diagrams shown in the second line of Fig.~\ref{fig:topod}
represent the color-favored, the gluon-annihilation, and the gluon-exchange penguin diagrams, respectively,
whose amplitudes will be denoted as $PT_{P(V)}$, $PE_{P(V)}$, and $PA_{P(V)}$, respectively.

Since a vector meson cannot be generated from the scalar or pseudoscalar operator, the amplitude $PT_P$ does not include contributions from the penguin operator $O_5$ or $O_6$. 
Consequently, the color-favored penguin amplitudes $PT_P$ and $PT_V$ can be expressed as
\begin{equation}\label{eq:PTP}
PT_P=-\frac{G_F}{\sqrt2}\lambda_ba_4(\mu)f_Vm_VF_1^{D\to P}(m_V^2)2(\varepsilon^*\cdot p_D),
\end{equation}
and
\begin{equation}\label{eq:PTV}
PT_V=-\frac{G_F}{\sqrt2}\lambda_b\left[a_4(\mu)-r_\chi a_6(\mu)\right]f_Pm_VA_0^{D\to V}(m_P^2)2(\varepsilon^*\cdot p_D),
\end{equation}
respectively, where $\lambda_b=V_{ub}V_{cb}^*$ with $V_{ub}$ and $V_{cb}^*$ being the CKM matrix elements, $a_{4,6}(\mu)=c_{4,6}(\mu)+c_{3,5}(\mu)/N_c$, with $c_{3,4,5,6}$ the Wilson coefficients, 
$r_\chi$ is a chiral factor, which takes the form 
\begin{equation}
r_\chi=\frac{2m_P^2}{(m_u+m_q)(m_q+m_c)},
\end{equation}
with $m_{u(c,q)}$ being the masse of $u(c,q)$ quark.
Note that the quark-loop corrections and the chromomagnetic-penguin contribution are also absorbed into $c_{3,4,5,6}$ as is shown in Ref.~\cite{Li:2013xsa}.

Similar to the amplitudes $E_{P,V}$, the amplitudes $PE$ only include the non-factorizable contributions as well. Therefore, the amplitudes $PE_{P,V}$, which are dominated by $O_4$ and $O_6$~\cite{Li:2012cfa}, can be parameterized as
\begin{equation}
PE_{P,V}^q=-\frac{G_F}{\sqrt2}\lambda_b\left[c_4(\mu)-c_6(\mu)\right]\chi_{q}^Ee^{i\phi_{q}^E}f_Dm_D\frac{f_Pf_V}{f_\pi f_\rho}(\varepsilon^*\cdot p_D). \label{eq:PE}
\end{equation}

For the amplitudes $PA_{P}$ and $PA_{V}$, the helicity suppression does not apply to the matrix elements of $O_{5,6}$, so the factorizable contributions exist. In the pole resonance model~\cite{Fusheng:2011tw}, after applying the Fierz transformation and the factorization hypothesis, the amplitudes $PA_{P}$ and $PA_{V}$ can be expressed as
\begin{equation}
\begin{split}
PA_P^q=-\frac{G_F}{\sqrt2}\lambda_b\Bigl[&(-2)a_6(\mu)(2g_S)\frac{1}{m_D^2-m_{P^*}^2}(f_{P^*}m_{P^*}^0)(f_D\frac{m_D^2}{m_c})\\
&+c_3(\mu)\chi_{q}^Ae^{i\phi_{q}^A}f_Dm_D\frac{f_Pf_V}{f_\pi f_\rho}\Bigr](\varepsilon^*\cdot p_D),
\end{split}\label{eq:PAP}
\end{equation}
and
\begin{equation}
\begin{split}
PA_V^q=-\frac{G_F}{\sqrt2}\lambda_b\Bigl[&(-2)a_6(\mu)(-2g_S)\frac{1}{m_D^2-m_{P^*}^2}(f_{P^*}m_{P^*}^0)(f_D\frac{m_D^2}{m_c})\\
&+c_3(\mu)\chi_{q}^Ae^{i\phi_{q}^A}f_Dm_D\frac{f_Pf_V}{f_\pi f_\rho}\Bigr](\varepsilon^*\cdot p_D),
\end{split}\label{eq:PAV}
\end{equation}
respectively, where $g_S$ is an effective strong coupling constant obtained from strong decays, e.g., $\rho \to \pi \pi$, $K^* \to K\pi$, and $\phi \to KK$, etc, and is set $g_S=4.5$~\cite{Fusheng:2011tw} in this work, $m_{P^*}$ and $f_{P^*}$ are the mass and decay constant of the pole resonant pseudoscalar meson $P^*$, respectively, and $\chi^A_q$ and $\phi^A_q$ are the strengths and the strong phases of the corresponding amplitudes.

From Fig.~\ref{fig:topod}, the decay amplitudes of $D^0  \to K^+K^*(892)^-$ and $D^0  \to K^-K^*(892)^+ $ in the FAT approach can be easily written down
\begin{equation}
\mathcal{M}^\lambda_{D^0\to K^+ K^{*-}}=T_{K^{*-}}+E^u_{K^+}+PT_{K^{*-}}+PE_{K^{*-}}^s+PE_{K^+}^u+PA_{K^{*-}}^s,\label{eq:MKp}
\end{equation}
and
\begin{equation}
\mathcal{M}^\lambda_{D^0\to K^- K^{*+}}=T_{K^-}+E^u_{K^{*+}}+PT_{K^-}+PE_{K^-}^s+PE_{K^{*+}}^u+PA_{K^-}^s,\label{eq:MKn}
\end{equation}
respectively, where $\lambda$ is the helicity of the polarization vector $\varepsilon(p,\lambda)$. In the FAT approach, the fitted non-perturbative parameters, $ \chi^E_{q,s} $, $ \phi^E_{q,s} $, $ \chi^A_{q,s} $, $ \phi^A_{q,s} $, are assumed to be universal, and can be determined by the data~\cite{Li:2013xsa}. 

\begin{table}[h]
\caption{\label{tab:penguin-contribution}
The magnitude of tree and penguin contributions (in unit of $10^{-3}$) corresponding the topological amplitudes in Eqs.~(\ref{eq:MKp}) and~(\ref{eq:MKn}). The factors `$\frac{G_F}{\sqrt{2}}\lambda_s(\varepsilon^*\cdot p_D)$' and `$-\frac{G_F}{\sqrt{2}}\lambda_b(\varepsilon^*\cdot p_D)$' are omitted in this table.
}
\begin{ruledtabular}
\begin{tabular}{c|cccccc}
\textrm{Decay modes}&
\textrm{$T_{K^{*-}}$}&
\textrm{$E^u_{K^+}$}&
\textrm{$PT_{K^{*-}}$}&
\textrm{$PE_{K^{*-}}^s$}&
\textrm{$PE_{K^+}^u$}&
\textrm{$PA_{K^{*-}}^s$}\\
\colrule
$D^0 \to K^+K^*(892)^-$ & $0.23$ & $-0.02+0.15\mathrm{i}$ & $3.83+4.32\mathrm{i}$ & $0.96-0.03\mathrm{i}$ & $0.13-0.81\mathrm{i}$ & $6.73+8.22\mathrm{i}$\\
\hline
 &
\textrm{$T_{K^-}$}&
\textrm{$E^u_{K^{*+}}$}&
\textrm{$PT_{K^-}$}&
\textrm{$PE_{K^-}^s$}&
\textrm{$PE_{K^{*+}}^u$}&
\textrm{$PA_{K^-}^s$}\\
\hline
$D^0 \to K^-K^*(892)^+$ & $0.44$ & $-0.02+0.15\mathrm{i}$ &$-23.3-19.3\mathrm{i}$ & $0.96-0.03\mathrm{i}$ & $0.13-0.81\mathrm{i}$ & $-8.53-5.53\mathrm{i}$\\
\end{tabular}
\end{ruledtabular}
\end{table}

In Table~\ref{tab:penguin-contribution}, we list the magnitude of each topological amplitudes for $D^0  \to K^+K^*(892)^-$ and $D^0  \to K^-K^*(892)^+ $ by using the global fitted parameters for $ D \to PV $ in Ref.~\cite{Li:2013xsa}. One can see from Table~\ref{tab:penguin-contribution} that the penguin contributions are greatly suppressed. $PT$ is dominant in the penguin contributions of $D^0 \to K^-K^*(892)^+$. While $PT$ is small in $D^0 \to K^+K^*(892)^-$, which is even smaller than the amplitude $PA$. This difference is because of the chirally-enhanced factor contained in Eq.~(\ref{eq:PTV}) while not in Eq.~(\ref{eq:PTP}). The very small $PE$ do not receive the contributions from the quark-loop and chromomagnetic penguins, since these two contributions to $c_4$ and $c_6$ are canceled with each other in Eq.~(\ref{eq:PE}). Besides, the relations $PE_V^s=PE_P^s$, $PE_V^u=PE_P^u$, and $PE_V^s\neq PE_V^u$ can be read from Table~\ref{tab:penguin-contribution}, this is because that the isospin symmetry and the flavor $SU(3)$ breaking effect have been considered.

\begin{table}[h]
\caption{\label{tab:BrFAT}
Branching ratios (in unit of $10^{-3}$) of singly-Cabibbo suppressed decays $D^0 \to K^+K^*(892)^-$ and $D^0 \to K^-K^*(892)^+$. Both experimental data~\cite{Cawlfield:2006hm,Aubert:2007dc,Olive:2016xmw} and theoretical predictions of FAT approach of the branching ratios are listed.
}
\begin{tabular}{p{2.8cm}<{\centering}|p{5cm}<{\centering}|p{5cm}<{\centering}}
\hline
\hline
\textrm{Form factors}&
\textrm{Br($D^0 \to K^+K^*(892)^-$)}&
\textrm{Br($D^0 \to K^-K^*(892)^+$)}\\
\hline
Pole & $1.57 \pm 0.04$ & $3.73 \pm 0.17$\\

Dipole & $1.69 \pm 0.04$ & $4.02 \pm 0.19$\\

CLF & $1.45 \pm 0.04$ & $4.44 \pm 0.20$\\

\hline
Exp. & $1.56\pm0.12$ & $4.38\pm0.21$\\
\hline
\hline
\end{tabular}
\end{table}

Since the form factors are inevitably model-dependent, we list in Table~\ref{tab:BrFAT} that the branching ratios of $D^0\to K^+K^*(892)^-$ and $D^0\to K^-K^*(892)^+$ predicted by the FAT approach, by various form factor models. The pole, dipole and covariant light-front (CLF) models are adopted. The uncertainties in Table~\ref{tab:BrFAT} mainly come from decay constants. The CLF model agrees well with the data for both decay channels, and other models are also consistent with the data. However, the model-dependence of form factor leads to large uncertainty of the branching fraction, as large as $20\%$. Because of the smallness of the Wilson coefficients and the CKM-suppression of the penguin amplitudes, the branching ratios are dominated by the tree amplitudes. Therefore, there is no much difference for the branching ratios whether we consider  the penguin amplitudes or not. 

\section{{\boldmath $\mathit{CP}$ asymmetries for $D^0 \to K^\pm K^*(892)^\mp$ and $D^0 \to K^+ K^- \pi^0$}  \label{sec:CP}}

The direct $\mathit{CP}$ asymmetry for the two-body decay $D\to PV$ is defined as
%%%%%%
\begin{equation}
A_{CP}^{D\to PV}=\frac{\left|\mathcal{M}_{D\to PV}\right|^2
-\left|\mathcal{M}_{\bar{D} \to \bar{P} \bar{V}}\right|^2}
{\left|\mathcal{M}_{D\to PV}\right|^2
+\left|\mathcal{M}_{\bar{D} \to \bar{P} \bar{V}}\right|^2}, \label{eq:diff-CP-a}
\end{equation}
%%%%%%
where $\mathcal{M}_{\bar{D}\to \bar{P} \bar{V}}$ represents the decay amplitude of the $\mathit{CP}$ conjugate process $\bar D \to \bar P \bar V$, such as $\bar{D}^0 \to  K^+ K^*(892)^-$ or $\bar{D}^0 \to  K^- K^*(892)^+$. In the framework of FAT approach, we predict very small direct $\mathit{CP}$ asymmetries of $D^0 \to K^+K^*(892)^-$ and $D^0 \to K^-K^*(892)^+$ presented in Table~\ref{tab:cpFAT}. The uncertainties induced by the model-dependence of form factor to the $\mathit{CP}$ asymmetries of $D^0 \to K^+K^*(892)^-$ and $D^0 \to K^-K^*(892)^+$, are about $30\%$ and $10\%$, respectively.
\begin{table}[h]
	\caption{\label{tab:cpFAT}
		$\mathit{CP}$ asymmetries (in unit of $10^{-5}$) of $D^0 \to K^+K^*(892)^-$ and $D^0 \to K^-K^*(892)^+$ predicted by the FAT approach with pole, dipole and CLF models adopted. The uncertainties in this table are mainly from decay constants.
	}
	\begin{tabular}{p{2.8cm}<{\centering}|p{5cm}<{\centering}|p{5cm}<{\centering}}
		\hline
		\hline
		\textrm{Form factors}&
		\textrm{$A_{CP}(D^0 \to K^+K^*(892)^-)$}&
		\textrm{$A_{CP}(D^0 \to K^-K^*(892)^+)$}\\
		\hline
		Pole & $-1.45 \pm 0.25$ & $3.60 \pm 0.23$\\
		
		Dipole & $-1.63 \pm 0.26$ & $3.70 \pm 0.24$\\
		
		CLF & $-1.27 \pm 0.25$ & $3.86 \pm 0.26$\\
		
		\hline
		\hline
	\end{tabular}
\end{table}

%\begin{equation}
%\begin{split}
%A_{CP}^{D^0 \to K^+K^{*-}}&=(-1.45 \pm 0.25^{+0.18}_{-0.10})\times 10^{-5},\\
%A_{CP}^{D^0 \to K^-K^{*+}}&=(3.86 \pm 0.26 \pm 0.10) \times10^{-5},
%\end{split}\label{eq:cp2b}
%\end{equation}

The differential $\mathit{CP}$ asymmetry of the three-body decay $D^0 \to K^+K^-\pi^0$, which is a function of the invariant mass of $s_{\pi^0 K^+}$ and $s_{\pi^0 K^-}$, is defined as
%%%%%%
\begin{equation}
A_{CP}^{D^0 \to K^+K^-\pi^0}(s_{\pi^0 K^+},s_{\pi^0 K^-})=\frac{\left|\mathcal{M}_{D^0 \to K^+K^-\pi^0}\right|^2
-\left|\mathcal{M}_{\bar{D}^0 \to K^-K^+\pi^0}\right|^2}
{\left|\mathcal{M}_{D^0 \to K^+K^-\pi^0}\right|^2
+\left|\mathcal{M}_{\bar{D}^0 \to K^-K^+\pi^0}\right|^2}, %\label{}
\end{equation}
%%%%%%
where the invariant mass $s_{\pi^0 K^\pm}=(p_{\pi^0}+p_{K^\pm})^2$. As can be seen from Eq.~(\ref{eq:totampl}), the differential $\mathit{CP}$ asymmetry $A_{CP}^{D^0 \to K^+K^-\pi^0}$ depends on the relative strong phase $\delta$,
which is impossible to be calculated theoretically because of its non-perturbative origin. 
Despite of this, we can still acquire some information of this relative strong phase $ \delta $ from data. By using a Dalitz plot technique~\cite{Cawlfield:2006hm,Rosner:2003yk,Bediaga:2009tr}, the phase difference $ \delta^\mathrm{exp} $ between $ D^0 $ decays to $ K^+ K^*(892)^- $ and $ K^- K^*(892)^+ $ can be extracted from data. One should notice that $ \delta^\mathrm{exp} $ is not the same as the strong phase $ \delta $ defined in Eq.~(\ref{eq:totampl}).  The strong phase $ \delta $ is the relative phase between  the decay amplitudes of $ D^0 \to K^+ K^*(892)^- $ and $ D^0 \to K^- K^*(892)^+ $. One the other hand, the  phase $ \delta^\mathrm{exp} $ is defined through
\begin{equation}
\mathcal{M}_{D^0 \to K^+K^-\pi^0}=\left(|\mathcal{M}_{K^{*+}}|+e^{i\delta^\mathrm{exp}}|\mathcal{M}_{K^{*-}}|\right)e^{i\delta_{K^{*+}}}
\end{equation}
in the overlapped region of the phase space, where $ \delta_{K^{*\pm}} $  is the phase of the amplitude $ \mathcal{M}_{K^{*\pm}} $:
\begin{equation}
\mathcal{M}_{K^{*\pm}}=|\mathcal{M}_{K^{*\pm}}|e^{i\delta_{K^{*\pm}}}.
\end{equation}	
Therefore, neglecting the CKM suppressed penguin amplitudes, $ \delta^\mathrm{exp} $ and $ \delta $ can be related by
	\begin{equation}
	\delta^\mathrm{exp}-\delta \approx \delta^{K^{*-} K^+ }-\delta^{K^{*+} K^- } ,\label{eq:spr}
	\end{equation}
where  $ \delta^{ K^{*\mp} K^\pm}=\arg(T_{K^{*
\mp}}+E^u_{K^\pm}) $ are the phases in tree-level amplitudes of $ D^0 \to K^\pm K^*(892)^\mp $, and are equivalent to $ \delta_{K^{*\mp}} $ if the penguin amplitudes are neglected. With the relation of Eq.~(\ref{eq:spr}), and $\delta^\mathrm{exp}=-35.5^\circ\pm4.1^\circ$ measured by the BABAR Collaboration~\cite{Aubert:2007dc}, we have $ \delta\approx -51.85^\circ \pm 4.1^\circ $.

\begin{figure}[h]
\centering
\includegraphics[width=0.618\textwidth]{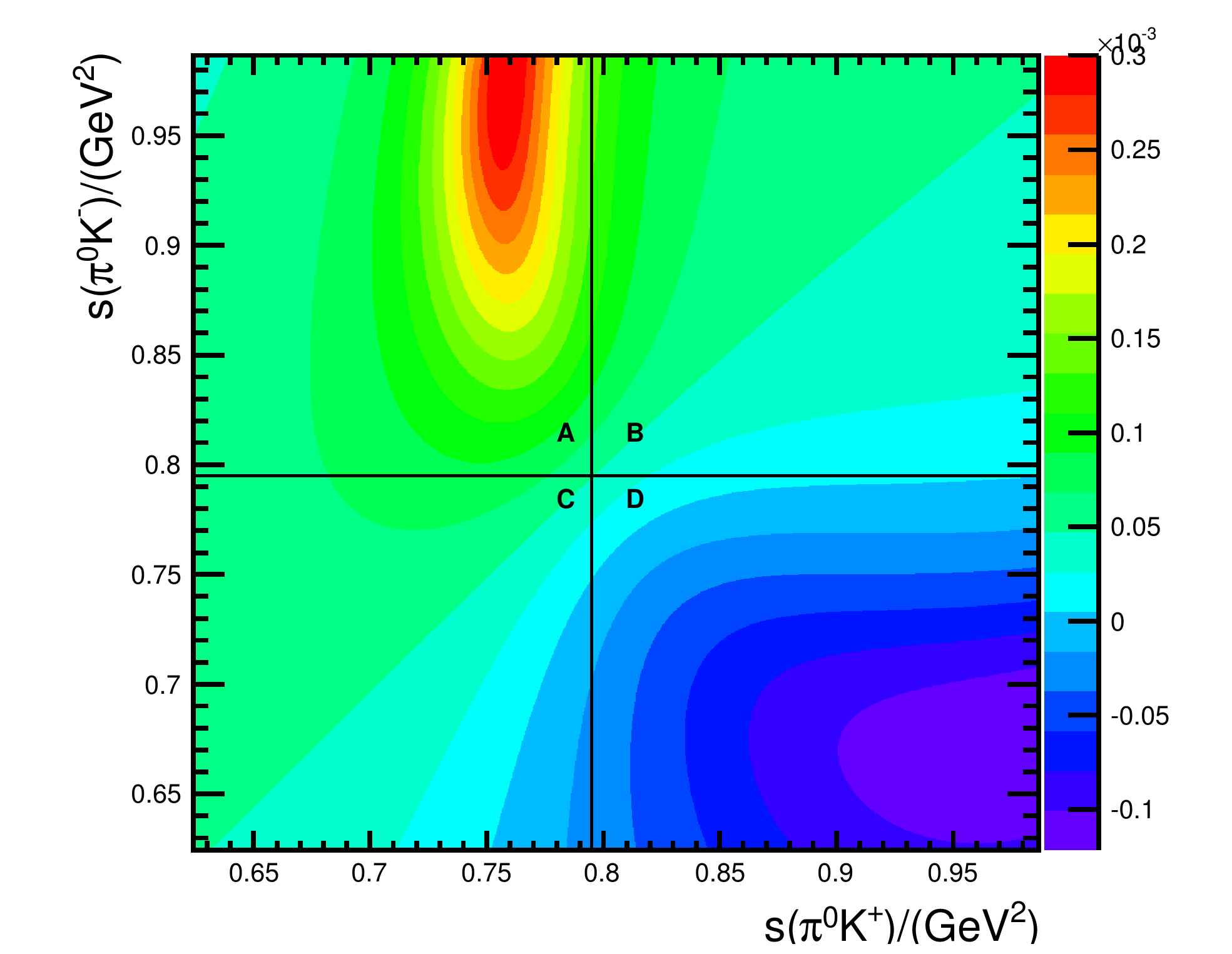}
\caption{The differential $\mathit{CP}$ asymmetry distribution of $D^0 \to K^+ K^- \pi^0$ in the overlapped region of $K^*(892)^-$ and $K^*(892)^+$ in the phase space.}
\label{fig:FATdiffAcp}
\end{figure}

In Fig.~\ref{fig:FATdiffAcp}, we present the differential $\mathit{CP}$ asymmetry of $ D^0 \to K^+ K^- \pi^0 $  in the overlapped region  of $K^*(892)^-$ and $K^*(892)^+$ in the phase space, with $\delta=-51.85^\circ$.
Namely,  we will focus on the region $m_{K^*}-2\Gamma_{K^*}<\sqrt{s_{\pi^0K^-}},\sqrt{s_{\pi^0K^+}}<m_{K^*}+2\Gamma_{K^*}$ of the phase space. 
One can see from Fig.~\ref{fig:FATdiffAcp} that the differential $\mathit{CP}$ asymmetry of $ D^0 \to K^+ K^- \pi^0 $ can reach  $ 3.0 \times 10^{-4} $ in the overlapped region, which is about 10 times larger than the $\mathit{CP}$ asymmetries of the corresponding two-body decay channels shown in Table~\ref{tab:cpFAT}. 

The behavior of the differential $\mathit{CP}$ asymmetry of $ D^0 \to K^+ K^- \pi^0 $ in Fig.~\ref{fig:FATdiffAcp} motivates us to separate this region into four areas, area A ($m_{K^*}<\sqrt{s_{\pi^0K^-}}<m_{K^*}+2\Gamma_{K^*}, m_{K^*}-2\Gamma_{K^*}<\sqrt{s_{\pi^0K^+}}<m_{K^*}$), area B ($m_{K^*}<\sqrt{s_{\pi^0K^-}}<m_{K^*}+2\Gamma_{K^*}, m_{K^*}<\sqrt{s_{\pi^0K^+}}<m_{K^*}+2\Gamma_{K^*}$), area C ($m_{K^*}-2\Gamma_{K^*}<\sqrt{s_{\pi^0K^-}}<m_{K^*}, m_{K^*}-2\Gamma_{K^*}<\sqrt{s_{\pi^0K^+}}<m_{K^*}$), and area D ($m_{K^*}-2\Gamma_{K^*}<\sqrt{s_{\pi^0K^-}}<m_{K^*}, m_{K^*}<\sqrt{s_{\pi^0K^+}}<m_{K^*}+2\Gamma_{K^*}$).
We further consider the observable of regional $\mathit{CP}$ asymmetry in areas A, B, C, D displayed in Table~\ref{tab:regionalcpFAT}, which is defined by
\begin{equation}
A_{CP}^\Omega=\frac{\int_\Omega(\left|\mathcal{M_\mathrm{tot}}\right|^2-\left|\mathcal{\overline{M}_\mathrm{tot}}\right|^2)\mathrm{d}s_{\pi^0K^-}s_{\pi^0K^+}}{\int_\Omega(\left|\mathcal{M_\mathrm{tot}}\right|^2+\left|\mathcal{\overline{M}_\mathrm{tot}}\right|^2)\mathrm{d}s_{\pi^0K^-}s_{\pi^0K^+}},
\end{equation}
where $\Omega$ represents a certain region of the phase space. 

\begin{table}[h]
\caption{\label{tab:regionalcpFAT}
Three from factor models: the pole, dipole and CLF models, are used for the regional $\mathit{CP}$ asymmetries (in unit of $10^{-4}$) in the four areas, A, B, C, D, of the phase space.
}
\begin{tabular}{p{3cm}<{\centering}|p{2.2cm}<{\centering}p{2.2cm}<{\centering}p{2.2cm}<{\centering}p{2.2cm}<{\centering}p{2.2cm}<{\centering}}
\hline
\hline
\textrm{Form factors}&
$A_{CP}^\mathrm{A}$   &
$A_{CP}^\mathrm{B}$   &
$A_{CP}^\mathrm{C}$   &
$A_{CP}^\mathrm{D}$   &
$A_{CP}^\mathrm{All}$  \\
\hline
Pole   & $0.87\pm0.11$ & $0.42\pm0.08$ & $0.39\pm0.07$ & $-0.30\pm0.08$ & $0.33\pm0.05$ \\

Dipole & $0.87\pm0.11$ & $0.41\pm0.08$ & $0.38\pm0.07$ & $-0.30\pm0.08$ & $0.32\pm0.05$ \\

CLF    & $0.84\pm0.10$ & $0.45\pm0.08$ & $0.42\pm0.07$ & $-0.25\pm0.08$ & $0.36\pm0.06$ \\
\hline
\hline
\end{tabular}
\end{table}
Comparing with the $\mathit{CP}$ asymmetries of two-body decays, the regional $\mathit{CP}$ asymmetries, from Table~\ref{tab:regionalcpFAT} are less sensitive to the models we have used. We would like to use only the CLF model for the following discussion. The uncertainties in Table~\ref{tab:regionalcpFAT} come from decay constants as well as the relative phase $\delta^\mathrm{exp}$. In addition, if we focus on the right part of area A, that is $m_{K^*}<\sqrt{s_{\pi^0K^-}}<m_{K^*}+2\Gamma_{K^*}, m_{K^*}-\Gamma_{K^*}<\sqrt{s_{\pi^0K^+}}<m_{K^*}$,  the regional $\mathit{CP}$ violation will be  $(1.09\pm0.16)\times10^{-4}$. 

The energy dependence of the propagator of the intermediate resonances can lead to a small correction to $\mathit{CP}$ asymmetry. For example, if we replace the Breit-Wigner propagator by the Flatt\'e Parametrization~\cite{Flatte:1976xu}, the correction to the regional $\mathit{CP}$ asymmetry will be about $1\%$.

Since the $\mathit{CP}$ asymmetry of $D^0\to K^+K^-\pi^0$ is extremely suppressed, it should be more sensitive to the NP. For example, some NPs have considerable impacts on the chromomagnetic dipole operator $ O_{8g} $~\cite{Golden:1989qx,Giudice:2012qq,Grossman:2006jg,Gronau:2014pra,Brod:2011re,Grossman:2012eb,Isidori:2011qw}. Consequently, the $\mathit{CP}$ violation in SCS decays may be further enhanced. 
In practice, the NP contributions can be absorbed into the corresponding effective Wilson coefficient $c_{8g}^\mathrm{eff}$~\cite{Beneke:2001ev,Li:2005kt}. 
For comparison, we first consider a relative small value of $c_{8g}^\mathrm{eff}$ (as in Ref.~\cite{Brod:2011re,Li:2012cfa}) lying within the range $(0,~1)$, the global $\mathit{CP}$ asymmetry of $D^0 \to K^*(892)^\pm K^\mp$ are no larger than $5\times10^{-5}$.    
Moreover, If we follow Ref.~\cite{Li:2013xsa} taking $c_{8g}^\mathrm{eff} \approx 10$ (While $c_{8g}^\mathrm{eff}=10$,  which is extracted from $\Delta A_\mathit{CP}$ measured by LHCb~\cite{Aaij:2011in}, is a quiet large quantity even for the coefficients corresponding tree-level operators. However, such large contribution can be realized if some NPs effects are pulled in. For example, the up squark-gluino loops in supersymmetry (SUSY) can arise significant contributions to $c_{8g}$. More details about the squark-gluino loops and other models in SUSY can be found in Ref.~\cite{Grossman:2006jg,Giudice:2012qq,Gabbiani:1996hi,Gabrielli:1995bd,Hagelin:1992tc}.), the global $\mathit{CP}$ asymmetries of $D^0 \to K^+ K^*(892)^- $ and $D^0 \to K^- K^*(892)^+ $ are then $(0.56\pm0.08)\times10^{-3}$ and $(-0.50\pm0.04)\times10^{-3}$, respectively.

\begin{figure}[h]
	\centering
	\subfigure[]{\label{fig:c8=1}
		\includegraphics[width=7.9cm]{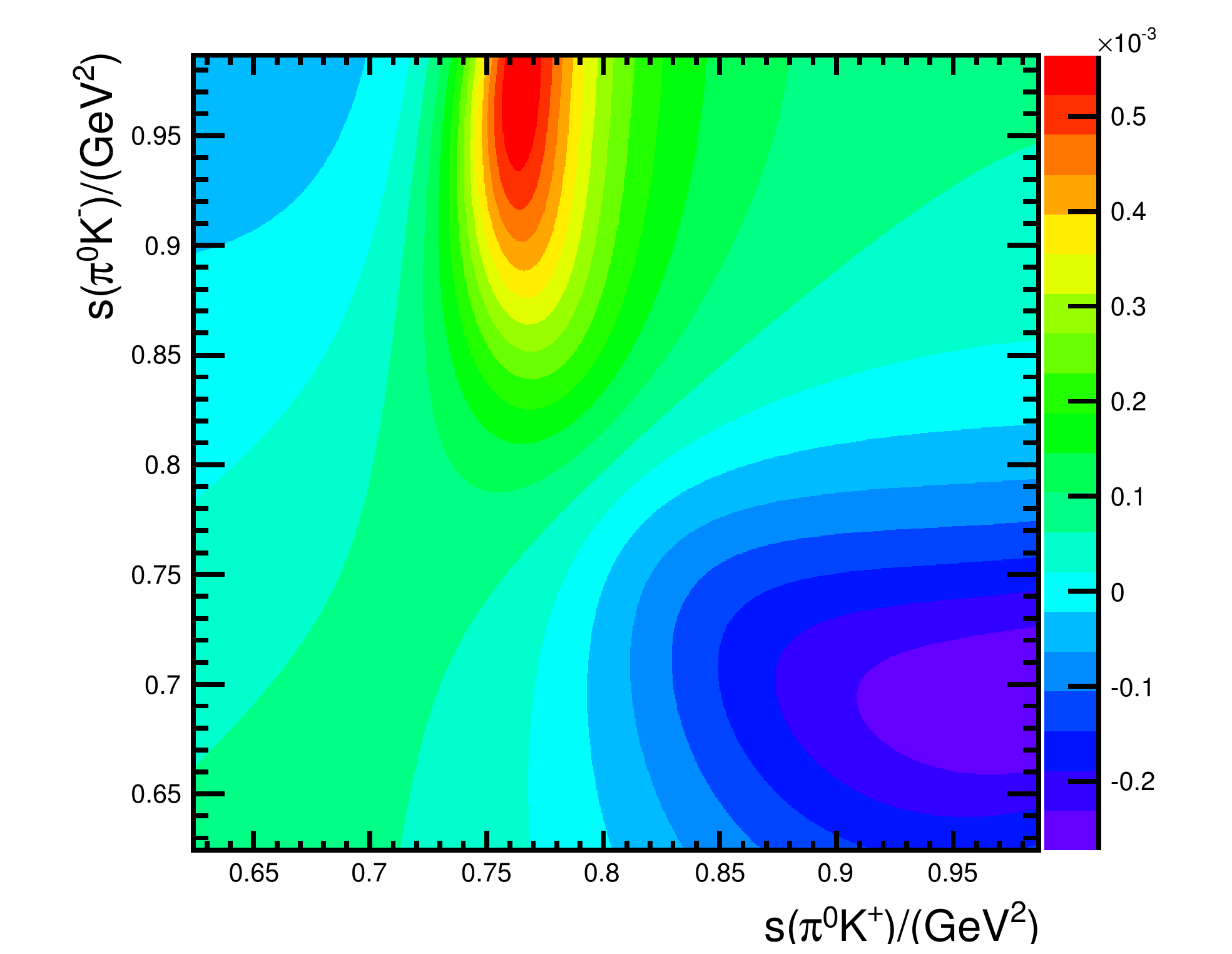}}
	\hspace{0.01in}
	\subfigure[]{\label{fig:c8=10}
		\includegraphics[width=7.9cm]{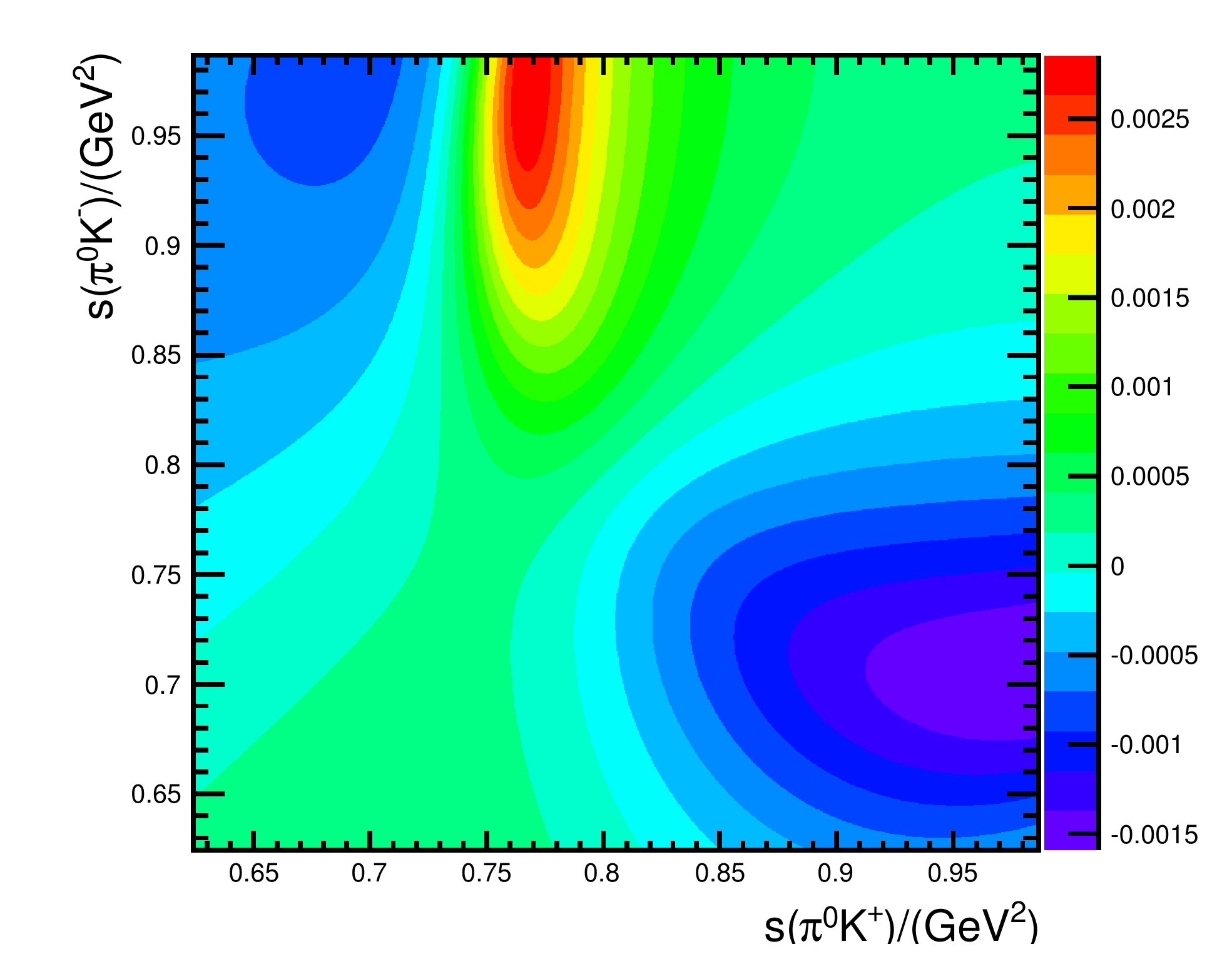}}
	\caption{The differential $\mathit{CP}$ asymmetry distribution of $D^0 \to K^+ K^- \pi^0$ for (a) $ c_{8g}^\mathrm{eff}=1 $ and (b) $ c_{8g}^\mathrm{eff}=10 $, in the overlapped region of $K^*(892)^-$ and $K^*(892)^+$ in the phase space.}
	\label{fig:NP}
\end{figure}

We further display the $\mathit{CP}$ asymmetry of $D^0 \to K^+K^-\pi^0$ in the overlapped region of $K^*(892)^-$ and $K^*(892)^+$ in Fig.~\ref{fig:c8=1} and Fig.~\ref{fig:c8=10} for $c_{8g}^\mathrm{eff}=1$ and $c_{8g}^\mathrm{eff}=10$, respectively. After taking the interference effect into account, the differential $\mathit{CP}$ asymmetry of $D^0 \to K^+K^-\pi^0$ can be increased as large as $5.5\times10^{-4}$ and $2.8\times10^{-3}$ for $c_{8g}^\mathrm{eff}=1$ and $c_{8g}^\mathrm{eff}=10$, respectively. The regional ones (in phase space of $ \sqrt{0.74}\  \mathrm{GeV}<\sqrt{s_{\pi^0K^-}}<\sqrt{0.81}\  \mathrm{GeV}, \sqrt{0.84}<\sqrt{s_{\pi^0K^+}}<m_{K^*}+2\Gamma_{K^*}$ ) can reach $(2.7\pm0.5)\times10^{-4}$ and $(1.3\pm0.3)\times10^{-3}$ for $c_{8g}^\mathrm{eff}=1$ and $c_{8g}^\mathrm{eff}=10$, respectively.

\section{discussion and conclusion\label{sec:dis-con}}
In this work, we studied $\mathit{CP}$ violations in $D^0 \to K^*(892)^\pm K^\mp \to K^+K^-\pi^0$ via the FAT approach.
The  $\mathit{CP}$ violations in two-body decay processes $D^0 \to  K^+ K^*(892)^-$ and $D^0 \to  K^- K^*(892)^+$ are very small, which are $(-1.27 \pm 0.25) \times 10^{-5}$ and $(3.86 \pm 0.26) \times 10^{-5}$, respectively. 
Our discussion shows that the $\mathit{CP}$ violation can be enhanced by the interference effect in three-body decay $D^0\to K^+K^-\pi^0$. The differential $\mathit{CP}$ asymmetry can reach $ 3.0\times 10^{-4} $ when the interference effect is taken into account. While the regional one can be as large as $ (1.09\pm0.16)\times 10^{-4} $. 
 
Besides, since the chromomagnetic dipole operator $O_{8g}$ are sensitive to some NPs, 
the inclusion of this kind of NPs will lead to a much larger global $\mathit{CP}$ asymmetries of $D^0 \to K^+ K^*(892)^- $ and $D^0 \to K^- K^*(892)^+ $, which are $(0.56\pm0.08)\times10^{-3}$ and $(-0.50\pm0.04)\times10^{-3}$, respectively. While the regional $\mathit{CP}$ asymmetry of $D^0\to K^+K^-\pi^0$ can be also increased to $(1.3\pm0.3)\times10^{-3}$ when considering the interference effect in the phase space. Since the $\mathcal{O}(10^{-3})$ of $\mathit{CP}$ asymmetry is attribute to the large $c_{8g}^\mathrm{eff}$, which is almost impossible for the SM to generate such large contribution, it will indicate NP if such $\mathit{CP}$ violation is observed. Here, we roughly estimate the number of $D^0\bar{D}^0$ needed for testing such kind of asymmetries, which is about $\frac{1}{Br}\frac{1}{A_\mathit{CP}^2} \sim 10^9$. This could be observed in the future experiments at Belle \uppercase\expandafter{\romannumeral2}~\cite{DePietro:2018crc,Abe:2010gxa}. While, the current largest $D^0\bar{D}^0$ yields is about $10^8$ at BABAR and Belle~\cite{Lees:2014ihu,Nisar:2015gvd}, and $10^7$ at BES\uppercase\expandafter{\romannumeral3} ~\cite{Ablikim:2015djc}.

\section{Conflicts of Interest}
The authors declare that there are no conflicts of interest regarding the publication of this paper.

\section{Acknowledegments}
This work was partially supported by National Natural Science Foundation of China (Project Nos. 11447021, 11575077, and 11705081), National Natural Science Foundation of Hunan Province (Project No. 2016JJ3104), the Innovation Group of Nuclear and Particle Physics in USC, and the China Scholarship Council.

%\bibliography{reference}
\appendix*
\section{Some useful formulas and input parameters\label{app:input}}

\subsection{Effective Hamiltonian and Wilson coefficients}
The weak effective Hamiltonian for SCS $D$ meson decays, based on the Operator Product Expansion (OPE) and Heavy Quark Effective Theory (HQET), can be expressed as~\cite{Buchalla:1995vs}

\begin{equation}
\mathcal{H}_{\mathrm{eff}}=\frac{G_F}{\sqrt{2}}
\left[
\sum_{q=d,s}\lambda_q(c_1O_1^q+c_2O_2^q)-\lambda_b(\sum_{i=3}^6c_iO_i+c_{8g}O_{8g})
\right]
+h.c.,
\end{equation}
where $G_F$ is the Fermi constant, $\lambda_q=V_{uq}V_{cq}^*$, $c_i (i=1,\cdots,6)$ is the Wilson coefficient, and $O_1^q$, $O_2^q$, $O_i (i=1,\cdots,6)$, and $O_{8g}$ are four-fermion operators which are constructed from different combinations of quark fields. The four-fermion operators take the following form
\begin{equation}
\begin{split}
O_1^q&=\bar u_\alpha\gamma_\mu(1-\gamma_5)q_\beta \bar q_\beta\gamma^\mu(1-\gamma_5)c_\alpha, \\
O_2^q&=\bar u\gamma_\mu(1-\gamma_5)q \bar q\gamma^\mu(1-\gamma_5)c, \\
O_3&=\bar u\gamma_\mu(1-\gamma_5)c\sum_{q'}\bar q'\gamma^\mu(1-\gamma_5)q', \\
O_4&=\bar u_\alpha\gamma_\mu(1-\gamma_5)c_\beta\sum_{q'}\bar q'_\beta\gamma^\mu(1-\gamma_5)q'_\alpha, \\
O_5&=\bar u\gamma_\mu(1-\gamma_5)c\sum_{q'}\bar q'\gamma^\mu(1+\gamma_5)q', \\
O_6&=\bar u_\alpha\gamma_\mu(1-\gamma_5)c_\beta\sum_{q'}\bar q'_\beta\gamma^\mu(1+\gamma_5)q'_\alpha, \\
O_{8g}&=-\frac{g_s}{8\pi^2}m_c\bar u\sigma_{\mu\nu}(1+\gamma_5)G^{\mu\nu}c,
\end{split}
\end{equation}
where $\alpha$ and $\beta$ are color indices and $q'=u, d, s$. Among all these operators, $O_1^q$ and $O_2^q$ are tree operators, $O_3-O_6$ are QCD penguin operators, and $O_{8g}$ is chromomagnetic dipole operator. The electroweak penguin operators are neglected in practice. One should notice that SCS decays receive contributions from all aforementioned operators while only tree operators can contribute to CF decays and DCS decays.

The Wilson coefficients used in this paper are evaluated at $\mu=1 \mathrm{GeV}$, which can be found in Ref.~\cite{Li:2012cfa}.

\subsection{CKM matrix}

We use the Wolfenstein parameterization for the CKM matrix elements, which up to order $\mathcal{O}(\lambda^8)$, read~\cite{Buras:1994ec,Charles:2004jd}
\begin{equation}
\begin{split}
V_{us}&=\lambda-\frac12A^2\lambda^7(\rho^2+\eta^2),  \\
V_{cs}&=1-\frac12\lambda^2-\frac18\lambda^4(1+4A^2)-\frac{1}{16}\lambda^6\left(1-4A^2+16A^2\left(\rho+i\eta\right)\right)\\
&\quad-\frac{1}{128}\lambda^8(5-8A^2+16A^4), \\
V_{ub}&=A\lambda^3(\rho-i\eta), \\
V_{cb}&=A\lambda^2-\frac12A^3\lambda^8(\rho^2+\eta^2),
\end{split}
\end{equation}
where $A, \rho, \eta$ and $\lambda$ are the Wolfenstein parameters, which satisfy following relation
\begin{equation}
\rho+i\eta=\frac{\sqrt{1-A^2\lambda^4}\left(\bar\rho+i\bar\eta\right)}{\sqrt{1-\lambda^2}\left[1-A^2\lambda^4\left(\bar\rho+i\bar\eta\right)\right]}.
\end{equation}
Numerical value of Wolfenstein parameters have been used in this work are as follows,
\begin{alignat}{2}
\lambda&=0.22548_{-0.00034}^{+0.00068}, \qquad &A&=0.810_{-0.024}^{+0.018}, \notag\\
\bar \rho&=0.145_{-0.007}^{+0.013}, &\bar \eta&=0.343_{-0.012}^{+0.011}.
\end{alignat}

\subsection{Decay constants and form factors}

In Eqs.~(\ref{eq:PAP}) and (\ref{eq:PAV}), the pole resonance model was employed for the matrix element $ \langle PV|\bar q_1 q_2|0\rangle $ in the annihilation diagrams. By considering  angular momentum conservation at weak vertex and all conservation laws are preserved at strong vertex, the matrix element $ \langle PV|\bar q_1 q_2|0\rangle $ is therefore dominated by a pseudoscalar resonance~\cite{Fusheng:2011tw},
\begin{equation}
\langle PV|\bar q_1 q_2|0\rangle=\langle PV|P^* \rangle\langle P^*|\bar q_1 q_2|0\rangle=g_{P^*PV}\frac{m_{P^*}}{m_D^2-m_{P^*}^2}f_{P^*},
\end{equation}
where $g_{P^*PV}$ is a strong coupling constant, $m_{P^*}$ and $f_{P^*}$ are the mass and decay constant of the pseudoscalar resonance $P^*$. Therefore,  $ \eta $ and $ \eta' $ being the dominant resonances for the final states of $ K^{*\pm}K^\mp $, which can be expressed as flavor mixing of $ \eta_q $ and $ \eta_s $,
\begin{equation}
\left(\begin{array}{c}
\eta \\
\eta'
\end{array}\right)=\left(\begin{array}{cc}	
\cos\phi & -\sin\phi \\
\sin\phi & \cos\phi
\end{array}\right)\left(\begin{array}{c}
\eta_q \\
\eta_s
\end{array}\right)
\end{equation}
where $ \phi $ is the mixing angle, $ \eta_q $ and $ \eta_s $ are defined by
\begin{equation}
\eta_q=\frac{1}{\sqrt{2}}(u \bar u+d \bar d), \quad \eta_s=s \bar s.
\end{equation}
The decay constants of $ \eta $ and $ \eta' $ are defined by
\begin{equation}
\begin{split}
\langle 0|\bar u \gamma_\mu\gamma_5 u|\eta(p)\rangle=i f_\eta^u p_\mu, \quad \langle 0|\bar u \gamma_\mu\gamma_5 u|\eta'(p)\rangle=i f_{\eta'}^u p_\mu, \\
\langle 0|\bar d \gamma_\mu\gamma_5 d|\eta(p)\rangle=i f_\eta^d p_\mu, \quad \langle 0|\bar d \gamma_\mu\gamma_5 d|\eta'(p)\rangle=i f_{\eta'}^d p_\mu, \\
\langle 0|\bar s \gamma_\mu\gamma_5 s|\eta(p)\rangle=i f_\eta^s p_\mu, \quad \langle 0|\bar s \gamma_\mu\gamma_5 s|\eta'(p)\rangle=i f_{\eta'}^s p_\mu,
\end{split}
\end{equation}
where 
\begin{equation}
f_\eta^u=f_\eta^d=\frac{1}{\sqrt{2}}f^q_\eta, \quad
f_{\eta'}^u=f_{\eta'}^d=\frac{1}{\sqrt{2}}f^q_{\eta'}.
\end{equation}
According to~\cite{Feldmann:1998vh,Feldmann:1998sh}, the decay constants of $ \eta $ and $ \eta' $ can be expressed as
\begin{alignat}{3}
f^q_\eta&=f_q\cos\phi, \quad &f^q_{\eta'}=f_q\sin\phi, \notag\\
f^s_\eta&=-f_s\sin\phi, \quad &f^s_{\eta'}=f_s\cos\phi.
\end{alignat}
where $ f_q=(1.07\pm0.02)f_\pi $ and $ f_s=(1.34\pm0.02)f_\pi $~\cite{Feldmann:1998vh}, the mixing angle $ \phi=(40.4\pm 0.6)^\circ $~\cite{Ambrosino:2009sc}. Other decay constants used in this paper are listed in Table~\ref{tab:decaycons}.

\begin{table}[h]
	\caption{\label{tab:decaycons}
		The meson decay constants used in this paper (MeV)~\cite{Olive:2016xmw,Ball:2006eu}. 
	}
	\begin{tabular}{p{2cm}<{\centering}p{2cm}<{\centering}p{2cm}<{\centering}p{2cm}<{\centering}p{2cm}<{\centering}}
		\hline
		\hline
		\textrm{$ f_{K^*} $}&
		\textrm{$ f_\rho $}&
		\textrm{$ f_K $}&
		\textrm{$ f_\pi $}&
		\textrm{$ f_D $}\\
		\hline
		
		$220 (5)$ & $ 216 (3) $ & $ 156 (0.4) $ & $ 130 (1.7) $ & $ 208 (10) $  \\
		
		\hline
		\hline
		
	\end{tabular}
\end{table}

The transition form factors $A_0^{D^0 \to K^{*-}}$ and $F_1^{D^0 \to K^-}$, based on the relativistic covariant light-front quark model~\cite{Cheng:2003sm}, are expressed as a momentum-dependent, 3-parameter form (the parameters can be found in Table~\ref{tab:formfactor})
\begin{equation}
F(q^2)=\frac{F(0)}{1-a(q^2/m_D^2)+b(q^2/m_D^2)^2}.\label{eq:ff}
\end{equation}
\begin{table}[h]
\caption{\label{tab:formfactor}
The parameters of $D\to K^*, K$ transitions form factors in Eq.~(\ref{eq:ff}).
}
\begin{tabular}{p{2.5cm}<{\centering}p{2.5cm}<{\centering}p{2.5cm}<{\centering}}
\hline
\hline
\textrm{Form factor}&
\textrm{$A_0^{D\to K^*}$}&
\textrm{$F_1^{D\to K}$}\\
\hline
$F(0)$ & 0.69 & 0.78  \\

$a$    & 1.04 & 1.05  \\

$b$    & 0.44 & 0.23  \\ 
\hline
\hline

\end{tabular}
\end{table}

\subsection{Decay rate}
The decay width takes the form
\begin{equation}
\Gamma_{D\to KK^*}=\frac{\left|\boldsymbol{p}_1\right|^3}{8\pi m_{K^*}^2}\left|\frac{\mathcal{M}_{D\to KK^*}}{\varepsilon^*\cdot p_D}\right|^2,
\end{equation}
where $\boldsymbol{p}_1$ represents the center of mass (c.m.) 3-momentum of each meson in the final state and is given by
\begin{equation}
\left|\boldsymbol{p}_1\right|=\frac{\sqrt{\left[(m_D^2-(m_{K^*}+m_K)^2)(m_D^2-(m_{K^*}-m_K)^2)\right]}}{2m_D},
\end{equation}
$\mathcal{M}$ is the corresponding decay amplitude.

\end{document}